# *Use of a nanoindentation fatigue test to characterize the ductile-brittle transition*

M. SKRZYPCZAK[1*], C. GUERRET-PIECOURT[1], S. BEC[1], J.-L. LOUBET[1], O. GUERRET[2]

*1 : **Ecole Centrale de Lyon – Laboratoire de Tribologie et Dynamique des Systèmes** – UMR 5513 CNRS/ECL/ENISE, 69134 Ecully, FRANCE*

*2 : **COATEX SAS**, 35 rue Ampère, 69730 Genay, FRANCE*

*\* : Corresponding author*

## Abstract

When considering grinding of minerals, scaling effect induces competition between plastic deformation and fracture in brittle solids. The competition can be sketched by a critical size of the material, which characterizes the ductile-brittle transition. A first approach using Vickers indentation gives a good approximation of the critical size through an extrapolation from the macroscopic to the microscopic scales. Nanoindentation tests confirm this experimental value. According to the grain size compared to the indent size, it can reasonably be said that the mode of damage is deformation-induced intragranular microfracture. This technique also enables to perform cyclic indentations to examine calcite fatigue resistance. Repeated loadings with a nanoindenter on $CaCO_3$ polycrystalline samples produce cumulative mechanical damage. It is also shown that the transition between ductile and brittle behaviour depends on the number of indentation cycles. The ductile domain can be reduced when the material is exposed to a fatigue process.

## 1. Introduction

Calcium carbonate ($CaCO_3$), also named calcite or marble in its crystalline form and chalk in its amorphous form, is widely used as filler in the industries of paper and painting because of its whiteness and its relative cheapness. Its use may require powders of fine granulometry which are generally obtained through ball milling [1]. In industry, experience shows that the smallest medium size obtained for calcite powder with this process is 1 or 2 micrometers, whatever the time the powder spends in the mill. Despite the number of studies dealing with powder size reduction [2-4], this limitation is still not clearly understood. Three phenomena are thought to be involved in this limitation:

1. Reagglomeration: during the grinding, small particles agglomerate, as they are more sensitive to attractive forces than bigger ones;
2. $CaCO_3$ grain size: if fracture was only intergranular, the limit size would be determined by the initial grain sizes of the bulk materials;
3. Intrinsic critical size: under this size, the energy balance becomes favorable to plastic deformations and no fracture can occur.

Many authors have already tried to define the reagglomeration causes, which origin can be electrostatic according to Uber [5] but can also be related to the capillarity forces according to Johnson [6] and to Balachandran. This author summarizes the theoretical order of magnitude of the different adhesion forces versus the particle size [7]. A complete and global overview is needed to fully understand this phenomenon and this is not the subject of this work.

To check if the second phenomenon, limitation by the initial grain size, is a possible explanation, polished samples were characterized by Electron Beam Scattering Diffraction (EBSD). The average grain size calculated from EBSD images was around 200 µm, which is consistent with data published by Cattaneo [8]. Consequently, it can be considered that both inter and intragranular fractures occur when grinding $CaCO_3$ powder with sizes below 200 µm, as it is needed by some industrial applications.

In this study, we focus on the third point, the determination of $CaCO_3$ critical size value and its evolution with cyclic loadings, for a given type of calcite, a Carrara marble. Although the ductile-brittle transition is well documented for metals [9], the transition between brittle and ductile zone in polycrystalline ceramics has been far less investigated, with only some recent studies [10-12]. To our knowledge, no study has been done yet concerning this transition for $CaCO_3$.

In a first part, in the aim to understand the brittle-ductile transition, the response of the material to a single indent has been studied. This has been first done by using Vickers indentation tests. From these results, a critical size for calcium carbonate could be defined by extrapolating the data. Indeed, due to the rather high applied load, cracks were always observed around Vickers indent and it was not possible to reach the ductile-only regime. That is why nanoindentation tests were then performed to obtain indents without crack in order to confirm the Vickers extrapolated value.

In a second part, since grinding induces a large number of shocks, the material response to cyclic stress has been characterized by multi-cycle nanoindentation tests in the ductile regime defined from the single cycle study. Cracks were observed for loads lower than the critical load determined for single indent which indicates that fatigue induced a shift of the ductile-brittle transition size with the number of cycles. The role of test parameters (loading speed, frequency, minimum unloading value in cyclic loading) on fracture propagation is briefly discussed.

## 2. Experimental methods

### 2.1. Samples

All tests were performed on freshly polished surfaces of calcium carbonate polycrystals ($CaCO_3$) from Carrara (Italy). Polishing was done up to a 1200 Grit grinding grade with a SiC abrasive disc (=P4000 grinding grade) to minimize the roughness of the samples. Polycrystals do not exhibit any specific crystallographic orientation according to EBSD images.

### 2.2. Single indent study

#### 2.2.1. Vickers indentation tests

Classical indentation techniques such as Vickers microhardness tests are geometrically limited by the four-sided pyramidal shape of the diamond as the tip defect is too large to permit a sufficient accuracy for small indentation loads. Although, some authors use Vickers indenter in nanoindentation tests for loads as low as 10 mN [13]. However, even if Vickers hardness tests are not designed for low load application, they remain useful to estimate an order of magnitude for the ductile-brittle transition in brittle materials.

During indentation, the energy given to the sample can be dissipated in two different ways: (i) a fracture occurs because the energy permits to create new surfaces or (ii) the material deforms plastically, the energy being dissipated by atoms rearrangement. Thermal effects also exist but are here neglected. The predominance of one or another phenomenon can be characterized by a size, referred here as the *critical breakage size* $a_{crit}$. This critical size results from the energetic balance between plastic deformation and crack propagation:

The energy $W_s$ needed to deform plastically a particle of volume $a^3$ is:
$$W_S = \frac{\sigma_p^2}{2E} a^3 \quad \text{(Eqn. 1)}$$
where $\sigma_p$ is the material yield stress and $E$ its Young modulus

On the other side, the energy $W_P$ needed to create a new surface $a^2$ is:
$$W_P = G a^2 \quad \text{(Eqn. 2)}$$
where $G$ is the crack propagation energy [14].

From these equations, the energetic balance becomes favorable to plastic deformation as soon as the size $a$ of the particle decreases under a critical breakage size named $a_{crit}$. This size $a_{crit}$ can be defined by:

$$a_{crit} = \frac{2GE}{\sigma_p^2}$$ (Eqn. 3)

Experimentally, brittle material heterogeneity leads to high scattering on $G$ values. Moreover, implementation of a characterization method to determine $\sigma_p$ is difficult. Consequently, accurate calculation of $a_{crit}$ using equation 3 is tricky.

However, the breakage critical size can be deduced by observing the transition between the plastic deformation and the creation of cracks when performing indentations. Thanks to microscopy techniques, such as optical microscopy or atomic force microscopy (AFM), indents on brittle materials can be geometrically characterized by the dimension $d$ of the plastically deformed area and the length $2c$ of the cracks around the indent, as defined in Figure 1.a

- The diagonal length $d$ of the residual impression is related to the resistance of the material to irreversible deformation and is linked to the material hardness $H$ via the relation:
  $$H = 1.854 \frac{P}{d^2}$$ (Eqn. 4)
  where $P$ is the applied load [15]

- The length of the cracks $2c$ is related to the material resistance to crack propagation. Calculation of the mode I stress intensity factor $K_{Ic}$ (in MPa.m$^{-1/2}$) can be done using [16]:
  $$K_{Ic} = 0.016 \left(\frac{E}{H}\right)^{1/2} \frac{P}{c^{3/2}}$$ (Eqn. 5).

Assuming that $E$, $H$ and $K_{Ic}$ are independent from the applied load $P$, the breakage critical size can be estimated from Vickers indentation observations.

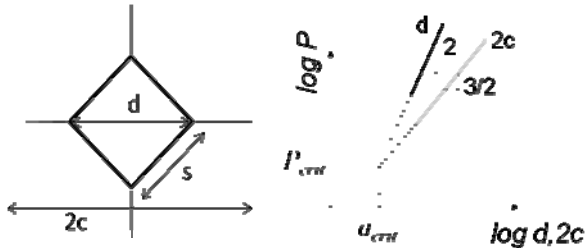

**Figure 1: a) Scheme of a Vickers indent with radial cracks; b) Theoretical evolution of diagonal length $d$ of the Vickers indentation and crack length $2c$ versus the applied load $P$**

When plotting the lengths $d$ and $2c$ versus the applied load $P$ from equation 4 and 5 respectively, straight lines are expected with slopes of 2 and 3/2 respectively. Thus the intersection point obtained by extrapolating the two straight lines determines a critical load $P_{crit}$. For higher loads than $P_{crit}$, brittle cracks appear. For loads lower than $P_{crit}$, only plastic deformation is observed. The associated characteristic length $a_{crit}$ corresponds to the critical size below which no cracking occurs (Figure 1b).

Several indentations were done at 0.25, 0.5, 1, 2 and 5N loads. The diagonal lengths of the impressions and the lengths of the cracks around the indents were measured immediately after the test with a precision of ± 0.2 µm using an optical microscope at x 80 magnification (Olympus BX51M). Average values were then calculated.

### 2.2.2. Nanoindentation tests

Tests at lower loads (2-10 mN) were performed with a nanoindenter to investigate directly the transition between ductile and brittle behaviour. Nanoindentation tests were performed using a Berkovich tip. According to Tabor [17], Vickers and Berkovich tips lead to a comparable deformation because they involve the same contact area $A$, (referred as $A_v$ and $A_b$ respectively for Vickers and Berkovich indenter) at the same contact depth. Thus, the results obtained from Berkovich nanoindentation can be compared to the Vickers indentation using a representative plasticity length defined by $\sqrt{(2A)}$. In the case of Vickers indentation $\sqrt{(2A)} = d$. For Berkovich indentation, $A$ is defined by the following equation [18]:
$$A = 24.5h_c^2$$ (Eqn. 6)
where $h_c$ is the contact depth. It is continuously calculated from indentation depth as a function of the applied load $P$ during two successive quasi-static loading-unloading cycles at a constant strain rate (0.015 s$^{-1}$) [19].
A comparison between the maximum depths measured after the first and the second loading steps enables to define whether a crack occurred or not after the first loading. When the depth difference between the two maxima is larger than 20 nm, it is assumed that breakage has happened. Atomic Force Microscopy (AFM) observations of the residual indents confirmed this interpretation: when a significant difference was observed between the two following maxima, cracks were observed while there was no crack when the two maxima were quite superimposed. For instance, Figure 2 shows an example of an indentation curve (a) and its corresponding AFM observation (b) for a maximum load of 10 mN where cracks were observed.

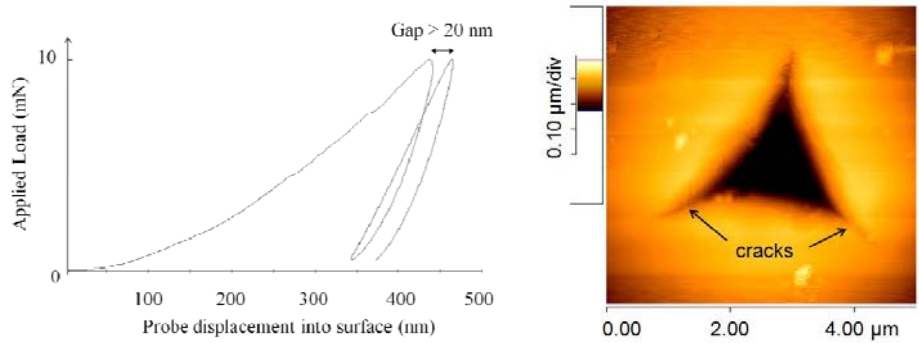

**Figure 2: a) Indentation curve on CaCO₃, b) Corresponding AFM image of the Berkovich indent with cracks (maximum load = 10 mN)**

Due to the low applied loads in nanoindentation, the indent size (a few micrometers) is significantly smaller than the grain size ($\approx$ 200 µm) so the tests were mainly performed on a single grain. This reduces the scattering of the results often observed when investigating brittle material mechanical properties. Furthermore, the small size of the indented zone enables to perform a large number of tests on the same sample. Such a statistical approach is necessary to cope with sample local heterogeneities as material defects are involved in crack initiation and propagation.

In this study, nanoindentation tests were performed at room temperature (25°C) and at relative humidity around 30%, using a MTS® Nanoindenter SA2. A Berkovich indenter was used and tests were done with loads $P$ of 1, 2, 2.5, 3, 4, 5, 6, 7.5 and 10 mN. For each load, 50 indents were done in order to reduce standard deviation of the measurement.

### 2.3. Fatigue tests

Fatigue tests consist in applying cyclic loading with a maximum load lower than the single cycle load for which radial cracks are observed. In this study, cyclic nanoindentation tests were conducted to investigate the influence of fatigue on the breakage critical size of CaCO₃. Each test was composed of 100 loading-unloading cycles. Indeed, in the case of ball mill grinding, the powder receives a large number of shocks and this can be considered as a fatigue process. Fatigue tests, especially through cyclic loading, have already been performed on marble [8], but these tests were performed at macroscale. Fatigue breakage process is well-known and has been described and explained by many authors. Wöhler first expressed the concept that a material can break when undergoing repeated stresses that are lower than its nominal stress[20]:

$$\sigma_{max} = C\hat{N}^{-1/n} \quad \text{(Eqn. 7)}$$

where $\hat{N}$ is the mean value of the number of cycles at failure $N_f$, $\sigma_{max}$ the maximum stress, $n$ the Weibull exponent and $C$ a constant.

If a power law is verified, a straight line is expected when plotting log ($\sigma_{max}$) versus log ($\hat{N}$). Its slope permits to calculate the Weibull exponent $n$, which characterizes the material resistance to fracture. This approach is generally valid for metals for a great number of cycles ($10^4$ to $10^8$ cycles) but recent studies confirm that it can be suitable for ceramics [21-24].

In this study, a fatigue test based on nanoindentation has been developed to characterize the material resistance to fracture under cyclic loading. For nanoindentation fatigue experiments, $\hat{N}$ corresponds to the average number of cycles at which cracks appear calculated from 50 tests. Nanoindentation was a particularly adapted tool for two reasons:
- Cyclic loads lower than the breakage critical load determined for CaCO₃ can be applied on the sample.
- As previously shown, the involved volume of material is small, which limits the scattering of the results according to the Griffith theory [25].

For each test, the number $\hat{N}$ of cycles needed to produce a crack in the material was recorded as a function of the maximum applied load $P$. The loading and unloading rate was set constant at 2.5 mN.s⁻¹. Thus the frequency of the cycles depends on the applied load, with a period varying between 2 and 8 seconds. According to Baïlon [26], change in frequency does not impact metals fracture toughness. In our case no significant difference on $\tilde{N}$ has been noticed in the range of the tested frequencies (0.125 to 0.5 Hz).

In order to ensure a correct location of all consecutive loadings at the same place, the unloading was not total. So during nanoindentation fatigue tests, the load varied between a maximum load $P_{max}$ and a minimum load $P_{min}$ that was kept constant at a value of 0.1 mN. This value is close enough to total unloading to permit crack propagation but may influence slightly the fatigue results [27].

Crack propagation was detected using the method employed by Beake et al., who compared the measured change in probe depth between two consecutive maximum loads to detect cracks [28-29]. In this study, we considered that a crack event leads to a depth increase at maximum loads larger or equal to 20 nm between two consecutive cycles. The indentation cycle that produces the crack is then the last one before this depth increase. The last cycle before the gap occurs correspond to the number of cycles $N_f$ needed to break the sample.

Figure 3 represents typical nanoindentation fatigue results: it clearly shows the change in depth during the test with an important displacement into surface (plastic deformations) during the $1^{st}$ load (1) followed by a slight slide into surface (<10 nm) before cracking (2) and a gap when cracking occur (3).

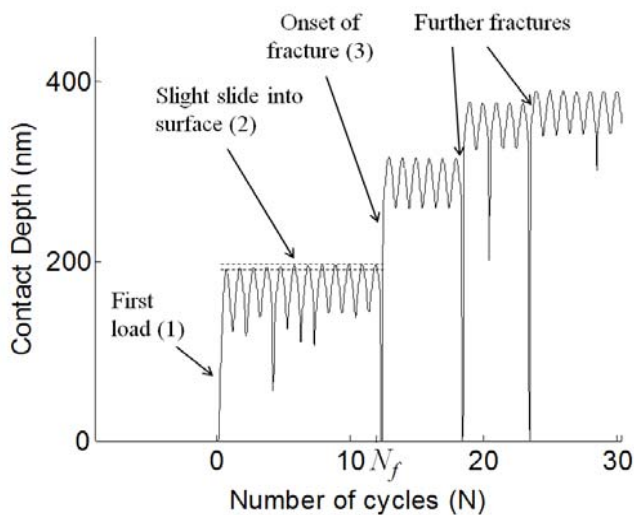

**Figure 3: Contact depth as a function of the number of cycles ($N$) for a maximum load = 2.5 mN (here cracking occurs for $N_f$ = 12). Further fractures are often observed after the first one occurred.**

# 3. Results

## 3.1. Critical load and size determined from Vickers Indentation

Figure 4 shows the evolution of the lengths $\sqrt{2A_v}$ and $2c$ as functions of the applied load. As expected, $\sqrt{2A_v}$ and $2c$ vary linearly versus the applied load $P$ and the slopes obtained are respectively 1.8 and 1.3 which is in rather good agreement with the expected values (2 and 1.5). The fact that Vickers indentation always produces cracks in the studied range may explain the observed difference between the theoretical and the experimental slopes for plasticity length.

The transition between plastic deformation and brittle fracture in $CaCO_3$ cannot be observed because the range of applied loads is higher than the ductile-brittle threshold of $CaCO_3$. Experimentally on $CaCO_3$, cracks are always observed around every Vickers indent. The transition size is indirectly obtained from extrapolation of the macroscopic results.

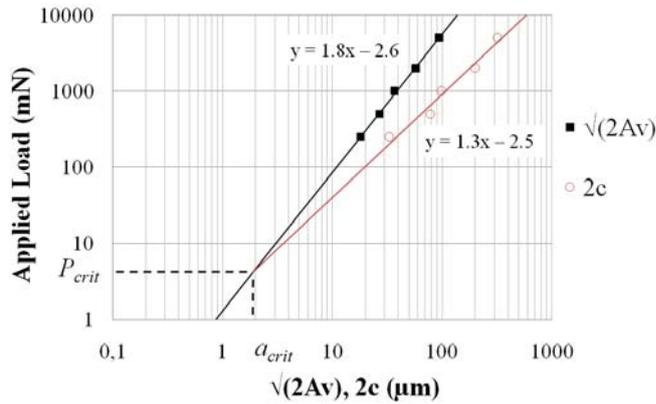

**Figure 4: Vickers indentation results: plasticity length $\sqrt{2A_v}$ (solid squares), fracture $2c$ (open circles). Extrapolations of these results give the critical size of $CaCO_3$**

By extrapolating the straight lines, the ductile-brittle critical size below which no cracking occurs is found to be around 1.9 µm. The corresponding load, i.e. the minimum load for which a crack should occur for a single cycle loading in the polycrystalline $CaCO_3$ sample, is between 4 and 5 mN.

### 3.2. Critical load from nanoindentation tests

In order to refine the previous results, nanoindentation tests were performed at loads chosen around the critical load obtained from the extrapolation of the Vickers indentation results. Using the depth increase criteria defined in 2.3, none of the 50 indentations made at loads up to 4 mN produces cracks after the first cycle whereas a non-negligible number of indentations performed at higher loads (6, 7.5 and 10 mN) produce cracks (Table 1). According to these results, the critical load for $CaCO_3$ is around 5 mN which is consistent with the previous Vickers results.

| Load (in mN) | 1 | 2 | 2.5 | 3 | 4 | 5 | 6 | 7.5 | 10 |
|---|---|---|---|---|---|---|---|---|---|
| % of cracked samples after 1 cycle | 0 | 0 | 0 | 0 | 0 | 2 | 14 | 24 | 25,5 |

**Table 1: Percentage of indentations for which a crack is detected after 1 cycle as a function of load**

For each load, the plasticity length $\sqrt{2A_b}$ for ductile behaviour is calculated from the contact depth for the maximum applied load using Eqn. 6. On Figure 5, $\sqrt{2A_b}$ is represented versus the applied load $P$ and compared to the Vickers dimensions ($\sqrt{2A_v}$ and $2c$). For the critical load $P$ = 5 mN, this calculation gives $\sqrt{2A_b} \approx 1.9\ \mu m$ which is in good agreement with $a_{crit}$ obtained from Vickers experiment. In the ductile zone (loads < 5 mN), the slope from the plasticity length points is 1.99. This value is closer to the theoretical one than the value obtained from Vickers indentation since nanoindentation enables to perform indents in the only-plastic regime.

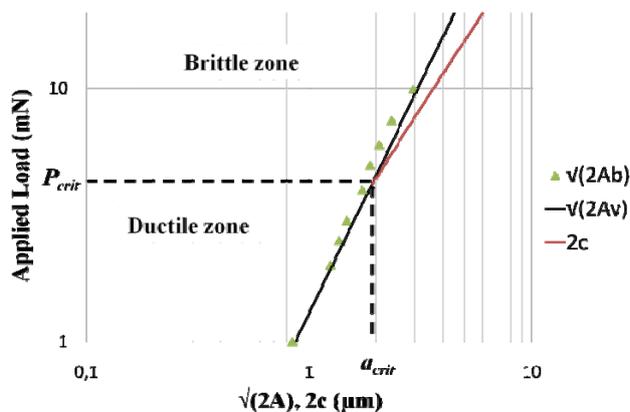

**Figure 5: Comparison between plastic deformation sizes induced by nanoindentation and Vickers indentation on $CaCO_3$ for single-cycle study**

### *3.3. Evolution of the brittle-ductile transition with fatigue*

For each maximum load (2-10 mN), 50 fatigue tests were performed and $\bar{N}$ corresponds to the average value of $N_f$. In Figure 6, $\bar{N}$ is plotted as a function of the applied load in a log-log plot (Wöhler diagram). A threshold value is observed for the applied load: for loads higher than or equal to 4 mN, $\bar{N}$ is almost constant at a low value (3-4 cycles). Thus, the fatigue results enable to determine with more accuracy the critical load of $CaCO_3$ which appears to be 4 mN instead of 5 mN as extrapolated with the single-shot study. Theoretically, $\bar{N}$ should be close to 1 for loads located in the brittle zone. One explanation could be the partial unloading during the nanoindentation procedure, which does not permit the crack propagation maybe due to partial stress relaxation.

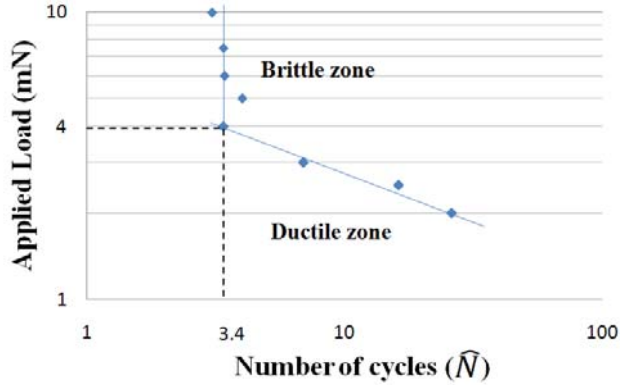

**Figure 6: Wöhler diagram of the contact fatigue results**

For P < 4 mN, a power law fits the fatigue results: $P = 5.8\bar{N}^{-0.32}$. According to Equation 7, the corresponding exponent value $n$ is 3.12. This value characterizes the material resistance to fracture. For high values of $n$, the material exhibits a good resistance to fatigue process. For comparison, Alumina F99.7, another brittle material used as technical ceramics because of its high resistance to fatigue, has a $n$ value of 6.55 [21].

The evolution of the ductile-brittle transition is plotted on Figure 7. It can be seen on this figure that the critical size of $CaCO_3$ is 1.7 µm for a single shock instead of 1.9 µm. When the material undergoes cyclic loadings, the ductile-brittle transition size decreases. For instance, this means that a particle with a size of 1.5 µm that should not be broken for a single shock (ductile behaviour according to Figure 5) could be broken after 7 shocks at 3 mN (Figure 6 and Figure 7).

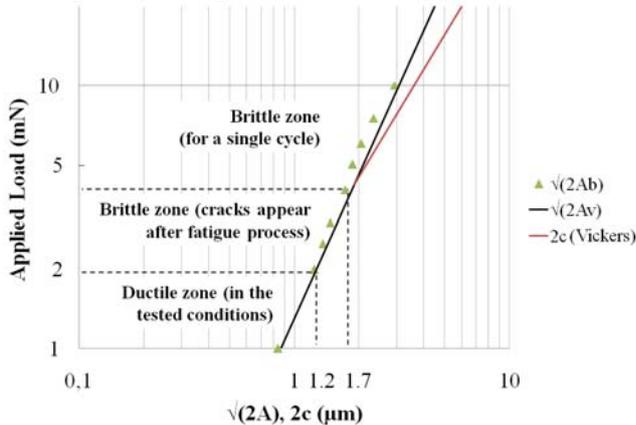

**Figure 7: Evolution of the brittle-ductile transition with fatigue**

## 4. Discussion

In this study, three main issues were investigated:

- the existence of a critical size between a ductile behaviour and a brittle behaviour and its determination from indentation tests;
- the behaviour under cyclic loading at microscale, which is close to classical macroscopic fatigue behaviour and can also be described by a power law in the investigated load range;
- the decrease of the transition size resulting from cyclic loadings.

## 4.1. Existence of a critical size and its determination

The existence of a critical size under which a brittle material become ductile is not a new concept [9-12]. Some authors even proposed to calculate this size using the material mechanical intrinsic properties. For instance, Kendall [30] proposed one formula based on material mechanical intrinsic properties (Vickers hardness $H_v$ and fracture toughness $K_{Ic}$):

$$a_{crit} \approx 72 * \left(\frac{K_{Ic}}{H_v}\right)^2 \quad \text{(Eqn. 8).}$$

This empiric formula was used by Wakeman to estimate the critical size of $CaCO_3$ [31]. The obtained value (1.19 μm for $CaCO_3$ in air) is in the order of magnitude of the measured one in this study. However, the scattering for $K_{Ic}$ observed in literature is great (from 2.4 to 5.2 MPa.m$^{1/2}$ for Carrara marble [32-33]). This large scatter for brittle materials makes the fracture toughness value not relevant to calculate the breakage critical size.

In the proposed method, this critical size was first experimentally extrapolated from Vickers indentation and then directly determined from nanoindentation tests. The possibility to apply very low loads on the sample using nanoindentation tests enables to observe the transition between ductile and brittle behaviours of $CaCO_3$ versus applied load. The contact depth was used to detect crack onset. Thus, $CaCO_3$ mechanical property values are not needed to estimate the critical size of the ductile-brittle transition. The critical size $a_{crit}$ was directly obtained.

## 4.2. Material behaviour under cyclic fatigue loadings

During grinding, particles undergo a high number of shocks. That is why understanding the material behaviour under fatigue loading may be a key point to improve its grinding. Fatigue is extensively studied for metals but less investigated for brittle materials [34-35]. Most studies are devoted to technical ceramics [21] or biomaterials [36]. Such ceramics have a good resistance to fatigue and fail for a high number of cycles, comparable to metals ($10^6$-$10^7$ cycles). To our knowledge only few papers [8, 37] were published concerning the fatigue behaviour of $CaCO_3$ at macroscale. Our study is focused on fatigue behaviour of $CaCO_3$ at microscale. The main difference between the two approaches is the characterized volume of material. At macroscale, several grains are involved which may introduce scattering due to grain boundaries. At microscale, tests are performed on single grains, which is the relevant size for grinding.

It was shown in 3.3 that, for loads lower than 4 mN, the oligo-cyclic fatigue results obtained for $CaCO_3$ can be described by a power law, classical for macroscopic fatigue. Thus, even a very brittle material, such as $CaCO_3$, can behave as a ductile material when tested under its critical load. As for ductile materials, it can be assumed that the successive loading-unloading cycles lead to residual stress accumulation that ends up to cracks propagation. The evolution of contact depth versus number of cycles shown in Figure 3 substantiates this hypothesis since no significant change in depth is observed during the fatigue process before the crack occurs. In the single indent study (2.2), a critical size $a_{crit}$ was defined. It corresponds to a threshold under which a particle receiving one shock could not be broken. According to micro and nanoindentation tests, this size $a_{crit}$ was estimated to be 1.9 μm and 1.7 μm respectively corresponding to a critical load $P_{crit}$ around 4 mN. Nanoindentation permits to investigate loads lower than $P_{crit}$ and thus the ductile zone. It was shown that, at such low loads, cracks propagation occurs when performing cyclic nanoindentation tests simulating a fatigue process. For instance, the threshold between ductile and brittle behaviour could thus be decreased to 1.2 μm for 2 mN cyclic loadings. In this study, the number of cycle performed on the sample was limited to 100 cycles. That is why crack occurrence was not significant for loads lower than 2 mN. By extrapolating the fatigue results obtained between 2 and 4 mN, the average number of cycles before fracture for a maximum applied load of 1 mN would be around 250 cycles.

For industrial applications, the fineness of the final ground powder, and consequently the decrease of the ductile-brittle threshold, is of major importance. Two ways can be explored to improve grinding efficiency. The first one is the increase of the number of shocks during comminution. According to the presented results, high number of shocks would lead to a lower ductile-brittle threshold and enhance fine particles creation. However, for small particles, the probability to receive a shock is smaller [2], which limits the efficiency of the fatigue process. In this work, this limit was not relevant since indentations were performed on a fixed sample. The shock probability will be taken into account in further work, through numerical simulations.

Decreasing the material resistance to fracture is another way to optimize grinding. Environmental parameters (hygrometry, temperature, pressure…) and/or chemical grinding aids are known to play a significant role in changing the material resistance to fracture [31]. A lower resistance to fracture would result in a decrease of the number of cycles needed to induce cracks for a given load. The Weibull exponent value $n$, defined earlier in this study, could give an order of magnitude of the change in the material resistance to fracture. The study of the role of grinding aids on the material resistance to fracture is in progress.

## Conclusion

This study has shown that single cycle nanoindentation tests enable to characterize the ductile-brittle transition of brittle materials. This transition can be observed directly instead of being deduced from extrapolated measurements at macroscopic scale. This method permits to determine the material critical size under which the energy balance becomes favourable to plastic deformations and no fracture can occur. A critical indentation load can be related to this critical size.

The behaviour under cyclic loading at microscale was investigated with nanoindentation tests which give the possibility to apply lower loads than the failure single cycle critical load. It is found that cyclic loading behaviour is close to usual macroscopic fatigue behaviour. In the tested range, it can be described by a classical fatigue power-law and plotted as Wöhler curve. Thus, failure can occur at lower load for high number of cycles and cracks can appear for particle sizes lower than the critical size defined for a single shock. This result can be related to the final powder granulometry obtained in CaCO$_3$ industrial milling process where particles smaller than the critical size are obtained.